\documentclass[reprint,twocolumn,superscriptaddress,showpacs,preprintnumbers,nofootinbib,aps,prd,
]{revtex4-1}

\usepackage{graphicx,color}
\usepackage{amsmath,amssymb,amsfonts}
\usepackage{stackrel}
\usepackage{enumitem}
\usepackage[english]{babel}
\usepackage{verbatim}
\usepackage{hyperref}
\usepackage{comment}
\usepackage{ulem}
\usepackage{xcolor}
\usepackage{subcaption}
\usepackage{dcolumn}
\usepackage{physics}
\usepackage{dsfont}

\definecolor{darkred}{rgb}{.7,.1,.1}
\usepackage{hyperref}
\hypersetup{
    pdfnewwindow=true,  
    colorlinks=true,     
    linkcolor=darkred,     
    citecolor=blue,      
    filecolor=blue,      
    urlcolor=blue        
}
 
\graphicspath{
  {./}
  {figures/}
}

\newcommand{\tf}{\texorpdfstring}
\newcommand{\gev}{~\text{GeV}}
\newcommand{\tev}{~\text{TeV}}

\newcommand{\onbb}{0\nu\beta\beta}

\def\nn{\nonumber}

\definecolor{orange}{rgb}{1,0.5,0}
\definecolor{amethyst}{rgb}{0.6, 0.4, 0.8}
\definecolor{antiquefuchsia}{rgb}{0.57, 0.36, 0.51}
\definecolor{byzantine}{rgb}{0.74, 0.2, 0.64}
\definecolor{blue-violet}{rgb}{0.54, 0.17, 0.89}
\definecolor{cadmiumred}{rgb}{0.89, 0.0, 0.13}
\definecolor{brightcerulean}{rgb}{0.11, 0.67, 0.84}

\begin{abstract}
\noindent

We revisit the contribution to the strong CP parameter $\bar \theta$ from leptonic CP violation at one-loop level in the minimal left-right symmetric model in the case of generalized parity as the left-right symmetry. The Hermitian neutrino Dirac mass matrix $M_D$ can be calculated using the light and heavy neutrino masses and mixings. We propose a parametrization of the right-handed neutrino mixing matrix $V_R$ and construct the heavy neutrino mass that maintains the Hermiticity of $M_D$. We further apply it to evaluate the one-loop $\bar\theta$, denoted as $\bar \theta_{loop}$, as a function of the sterile neutrino masses for explicit examples of $V_R$. By requiring the magnitude of $\bar \theta_{loop}\lesssim 10^{-10}$, we derive the upper limits on the sterile neutrino masses, which are within reach of direct searches at the Large Hadron Collider and neutrinoless double beta decay experiments. Furthermore, our parametrization is pertinent for other phenomenological investigations.

\end{abstract}

\begin{document}

\setlength{\abovedisplayskip}{6pt}
\setlength{\belowdisplayskip}{6pt}

\newcommand{\SUNYATSET}{\affiliation{School of Physics and Astronomy, Sun Yat-sen University, Zhuhai 519082, P.R. China.}}

\title{Calculable neutrino Dirac mass matrix and one-loop $\bar{\theta}$ in the minimal left-right symmetric model}


\author{Gang Li}
\email{ligang65@mail.sysu.edu.cn}
\SUNYATSET

\author{Ding-Yi Luo}
\email{luody25@mail2.sysu.edu.cn}
\SUNYATSET

\author{Xiang Zhao}
\email{zhaox88@mail2.sysu.edu.cn}
\SUNYATSET

\maketitle

\section{Introduction}
\label{introduction}

The standard model (SM) of particle physics has achieved great success. However, the origin of neutrino masses and the strong CP problem remains unsolved, serving as compelling motivations for physics beyond the SM (BSM). These two problems might have intrinsic connections, even though they appear in the weak and strong sectors at low energies. 

If neutrinos are Majorana fermions, they could acquire Majorana masses in the seesaw mechanism~\cite{Minkowski:1977sc,Gell-Mann:1979vob, Yanagida:1979as, Glashow:1979nm, Mohapatra:1979ia}, making them naturally small. In the type-I~\cite{Minkowski:1977sc, Gell-Mann:1979vob, Yanagida:1979as, Glashow:1979nm, Mohapatra:1979ia} and type-II~\cite{Konetschny:1977bn, Magg:1980ut, Schechter:1980gr, Mohapatra:1980yp, Lazarides:1980nt, Cheng:1980qt} seesaw mechanisms,  right-handed neutrinos and scalar triplet are introduced, respectively. In the minimal left-right symmetric model (MLRSM)~\cite{Pati:1974yy,Mohapatra:1974gc,Senjanovic:1975rk,Senjanovic:1978ev,Mohapatra:1979ia,Mohapatra:1980yp}, 
the neutrino masses can receive contributions from both type-I and type-II seesaw mechanisms $M_\nu = M_L - M_D^T M_N^{-1} M_D$ (cf. Eq.~\eqref{eq:seesaw}). 
In case of generalized parity or charge conjugation as the left-right symmetry~\cite{Senjanovic:2011zz}, dubbed case $\mathcal{P}$ or $\mathcal{C}$, respectively, the MLRSM is highly predictive, which has been extensively studied~\cite{Maiezza:2010ic,Tello:2010am,Nemevsek:2011aa,Bertolini:2014sua,Maiezza:2014ala,Senjanovic:2014pva,Cirigliano:2018yza,Zhang:2020lir,Li:2020flq,Dekens:2021bro,deVries:2022nyh,Maiezza:2021dui}. Moreover, it was found that in the MLRSM, one can calculate the neutrino Dirac mass matrix $M_D$ in terms of the light and heavy neutrino masses and mixings~\cite{Nemevsek:2012iq,Senjanovic:2016vxw,Senjanovic:2018xtu,Kiers:2022cyc,Kriewald:2024cgr}.
As a contrast, the expression of neutrino Dirac mass matrix $M_D$ in Casas-Ibarra parametrization~\cite{Casas:2001sr} in type-I seesaw models is still dependent on an arbitrary complex orthogonal matrix.

The strong CP problem is about the extremely small parameter $\bar \theta \lesssim 10^{-10}$~\cite{Crewther:1979pi,Abel:2020pzs,Graner:2016ses} that violates CP in the strong sector of the SM. The most popular solution to the strong CP problem is the Peccei-Quinn mechanism~\cite{Peccei:1977hh,Peccei:1977ur}, which leads to the existence of the axion~\cite{Weinberg:1977ma,Wilczek:1977pj} and has thus drawn a lot of theoretical attention~\cite{Dine:1981rt,Zhitnitsky:1980tq,Kim:1979if,Shifman:1979if,DiLuzio:2016sbl,Sun:2020iim,DiLuzio:2021pxd} as well as experimental interest~\cite{DiLuzio:2020wdo}. Additionally, the strong CP problem can also be addressed by imposing discrete symmetries~\cite{Mohapatra:1978fy,Babu:1988mw,Babu:1989rb,Barr:1991qx,Nelson:1983zb,Barr:1984qx}. Parity solutions to the strong CP problem in the left-right symmetric models were considered in Refs.~\cite{Mohapatra:1978fy,Babu:1988mw,Babu:1989rb}, and have been further studied recently~\cite{Bertolini:2019out,Ramsey-Musolf:2020ndm,Craig:2020bnv,Babu:2023srr}. In both SM and BSM scenarios, we can separate $\bar \theta = \theta +  \arg \det (M_u M_d)$, where $\theta$ is the coefficient of $G\tilde{G}$ term in the Lagrangian, and $\arg \det (M_u M_d)$ is included since the up-type and down-type quark mass matrices $M_u$ and $M_d$ are in general non-Hermitian~\cite{Hook:2018dlk}. 

In the MLRSM for the case $\mathcal{P}$, $\theta$ vanishes at tree level and $\bar \theta$ is equal to $\arg \det (M_u M_d)$. It has been shown that $\bar \theta \simeq \sin \alpha \tan (2\beta) m_t/(2m_b) $~\cite{Maiezza:2014ala,Bertolini:2019out}, where $\alpha$ and $\beta$ are defined in Eq.~\eqref{eq:bi-doublet}, $m_t$ and $m_b$ denote the masses of top and bottom quarks, respectively. Thus in order to satisfy the constraint from measurements of neutron electric dipole moments~\cite{Abel:2020pzs,Graner:2016ses,Crewther:1979pi} on $\bar \theta \lesssim 10^{-10}$, $\sin \alpha \tan (2\beta) \to 0$ is required.
Nevertheless, the MLRSM does not provide any reason why the phase $\alpha$ takes an extremely small value.
Thus there is no solution to the strong CP problem provided by the MLRSM, which has the minimal fermion content, contrasting with the left-right symmetric models discussed in Refs.~\cite{Mohapatra:1978fy,Babu:1988mw,Babu:1989rb,Craig:2020bnv,Babu:2023srr}.

However, even if the quark mass matrices are (nearly) Hermitian, 
leptonic CP violation would induce $\bar \theta$ at one-loop level,
which might exceed the aforementioned bound as pointed out in Ref.~\cite{Kuchimanchi:2014ota}. 
Instead of being a problem, Senjanovic et al.~\cite{Senjanovic:2020int} demonstrated that the one-loop $\bar \theta$ in the MLRSM implies an upper bound on the masses of sterile neutrinos, which is complementary to the direct searches at the Large Hadron Collider (LHC)~\cite{Nemevsek:2018bbt}. 
As obtained in Ref.~\cite{Senjanovic:2020int}, $\bar \theta_{loop}$ is proportional to $\Im \Tr (M_N^\dagger M_N [M_D, M_\ell])$, where $M_\ell$ denotes the charged lepton mass matrix, and the neutrino Dirac mass matrix $M_D$ is determined by the light and heavy neutrino masses and mixings.  However, it was shown that~\cite{deVries:2022nyh} $\bar \theta_{loop}$ might vanish in the type-I seesaw dominance scenario for specific benchmark choices of the right-handed neutrino mixing matrix $V_R$, which hindered the attempt to search for sterile neutrinos contributing to $\bar \theta$ with neutrinoless double beta ($\onbb$) decay~\cite{deVries:2022nyh}.

In this work, we propose a parametrization of right-handed neutrino mixing $V_R$ in the MLRSM for the case $\mathcal{P}$ and construct the heavy neutrino mass matrix $M_N$, for which the Hermiticity of the neutrino Dirac mass matrix $M_D$ is maintained. We then evaluate the one-loop $\bar\theta$ for the general seesaw relation for explicit examples of $V_R$,
and obtain non-vanishing $\bar \theta_{loop}$ as a function of the sterile neutrino masses. By using the bound $|\bar \theta_{loop}| \lesssim 10^{-10}$, we can then obtain the upper limits of the sterile neutrino masses. Our parametrization of $V_R$ is pertinent for other phenomenological studies, such as $\onbb$ decay and LHC searches~\footnote{The latter is in the vein of the methodology of Ref.~\cite{Solera:2023kwt}, which studied the impact of general textures of the right-handed quark mixing matrix without manifest left-right symmetry on the production of right-handed gauge boson at the LHC.
}.

The remainder of the paper is organized as follows. In the next section, we provide a brief introduction of the MLRSM for the case $\mathcal{P}$. Sec.~\ref{sec:cal-MD} delves into the calculation of neutrino Dirac mass matrix $M_D$ in the Senjanovic-Tello method, and the parametrization of $V_R$ and $M_N$.  In Sec.~\ref{sec:theta}, we evaluate the one-loop $\bar\theta$ for explicit examples of $V_R$. We conclude in Sec.~\ref{sec:conclusion}.

\section{Minimal left-right symmetric model}
\label{sec:MLRSM}

The MLRSM is based on the gauge group $SU(3)_c \times SU(2)_L \times SU(2)_R \times U(1)_{B-L}$, which was proposed to explain the origin of neutrino masses~\cite{Mohapatra:1979ia,Mohapatra:1980yp}. Three right-handed neutrinos $\nu_R$ and scalar triplets $\Delta_{L,R}$ are introduced
\begin{align}
\ell_{L,R} &= \begin{pmatrix}
    \nu\\
    e
\end{pmatrix}_{L,R},\
\Delta_{L,R} = 
\begin{pmatrix}
\delta_{L,R}^+/\sqrt{2} & \delta_{L,R}^{++}\\
\delta_{L,R}^0 & -\delta^+_{L,R}/\sqrt{2}
\end{pmatrix}\;,
\end{align}
where the flavor indices of leptons are omitted.
Besides, the scalar bi-doublet $\Phi$ exists, which is written as 
\begin{align}
\Phi = [\phi_1, i\sigma_2 \phi_2^*]\;,\quad 
\phi_i = \begin{pmatrix}
\phi_i^0\\
\phi_i^-
\end{pmatrix}\;,\quad i=1,2\;,
\end{align}
where $\sigma_2$ is the second Pauli matrix.
If generalized parity is taken as the left-right symmetry, i.e., case $\mathcal{P}$, we have
\begin{align}
\Delta_L \leftrightarrow \Delta_R\;, \quad
\Phi \leftrightarrow \Phi^\dagger\;.
\end{align}
The leptonic Yukawa interactions are
\begin{align}
\label{eq:lag-yuk}
\mathcal{L} &= -\bar\ell_L \left( Y_1 \Phi - Y_2 \sigma_2 \Phi^* \sigma_2 \right) \ell_R\nn\\
&\quad -\dfrac{1}{2} \left( \ell_L^T C Y_L i\sigma_2 \Delta_L \ell_L + \ell_L^T C Y_R i\sigma_2 \Delta_R \ell_R \right)\nn\\
&\quad + {\rm h.c.}\;,
\end{align}
where
$C = i\gamma^0 \gamma^2$ is the charge conjugation matrix,
h.c. denotes the Hermitian conjugate terms. The left-right symmetry is spontaneously broken once the right-handed triplet $\Delta_R$ develops a vacuum expectation value (vev), $v_R =\langle \delta_R^0\rangle$.
After the electroweak symmetry breaking, $\Phi$ develops vevs
\begin{align}
\label{eq:bi-doublet}
\langle \Phi \rangle = v {\rm diag}(c_\beta,-s_\beta e^{-i\alpha})
\end{align}
with $c_\beta \equiv \cos\beta$, $s_\beta \equiv \sin\beta$ and $v\simeq 174\gev$. Then the left-handed triplet $\Delta_L$ would get the vev $v_L$, which is generally complex~\cite{Zhang:2007da} and proportional to $v^2/v_R$~\cite{Mohapatra:1980yp,Deshpande:1990ip}. 
Defining $N_L = \nu_R^c$, 
one obtains the neutrino mass terms
\begin{align}
    \mathcal{L}_\nu = -\dfrac{1}{2} (\bar \nu_L^c, \bar N_L^c) M_n 
    \begin{pmatrix}
        \nu_L \\
        N_L
    \end{pmatrix}
    + {\rm h.c.}\;,
\end{align}
where the full neutrino mass matrix is defined as
\begin{align}
M_n \equiv
\begin{pmatrix}
M_L & M_D^T\\
M_D & M_N
\end{pmatrix}\;.
\end{align}
The neutrino Majorana and Dirac neutrino mass matrices are
\begin{align}
M_L &= \dfrac{v_L}{v_R} U_e^T M_N^* U_e^* \;,\\
M_N &= Y_R^* v_R\;,\\
M_D &= -v \left( Y_1 c_\beta + Y_2 s_\beta e^{-i\alpha}  \right)\;.
\end{align}
After block diagonalizing the neutrino mass matrix, we can obtain the light neutrino masses
\begin{align}
\label{eq:seesaw}
M_\nu = M_L - M_D^T \dfrac{1}{M_N} M_D\;,
\end{align}
which is a general seesaw relation including contributions from both type-I and type-II mechanisms. If $v_L$ is negligibly small, it is reduced to the type-I seesaw dominance scenario. 

As shown in Ref.~\cite{Senjanovic:2018xtu}, in the MLRSM for the case $\mathcal{P}$ we have
\begin{align}
    M_D - U_e M_D^\dagger U_e \propto s_\alpha t_{2\beta}\;,
\end{align}
where $s_\alpha \equiv \sin\alpha$, $t_{2\beta} \equiv \tan(2\beta)$, $U_e$ is the matrix that diagonalizes the charged lepton mass matrix. Thus, in the limit $s_\alpha t_{2\beta}\to 0$, $M_D$ is Hermitian and $U_e  = \pm \mathds{1}$.

\section{Calculable neutrino Dirac mass matrix}
\label{sec:cal-MD}

\subsection{Senjanovic-Tello method}

It has been shown by Senjanovic and Tello~\cite{Senjanovic:2016vxw,Senjanovic:2018xtu} that the neutrino Dirac mass matrix $M_D$ can be determined with the light and heavy neutrino masses and mixings in the limit $s_\alpha t_{2\beta}\to 0$. In the following, we will briefly introduce the general method they proposed in Ref.~\cite{Senjanovic:2018xtu}.

From $M_D = M_D^\dagger$, 
Eq.~\eqref{eq:seesaw} can be expressed as
\begin{align}
    HH^T=\frac{v_L}{v_R} \mathds{1}-
    \frac{1}{\sqrt{M_N}} M_\nu^* \frac{1}{\sqrt{M_N}}
    \label{eq:calculate HHT}\;,
\end{align}
where the Hermitian matrix $H$ is defined as 
\begin{align}
\label{eq:H-matrix}
    H=\frac{1}{\sqrt{M_N}} M_D \frac{1}{\sqrt{M_N}}\;.
\end{align}
One can then decompose $HH^T$ as 
\begin{align}
\label{eq:H-decompose}
    H H^T = O s O^T\;,
\end{align}
using the fact that $HH^T$ is symmetric. In the above, $O$ is a complex orthogonal matrix and $s$ is the symmetric normal form. The matrices $O$ and $s$ are obtained
from Eqs.~\eqref{eq:calculate HHT}~\eqref{eq:H-decompose}.
The matrix $H$ itself can be expressed as
\begin{align}
\label{eq:H-matrix2}
    H = O \sqrt{s} E O^\dagger\;.
\end{align}
with $E$ being determined by the Hermitian condition $H = H^\dagger$.

\begin{align}
    \sqrt{s} E = E \sqrt{s^*}\;,\quad
    E^T = E^* = E^{-1}\;.
\end{align}

Comparing Eq.~\eqref{eq:H-matrix2} with Eq.~\eqref{eq:H-matrix}, one readily get 
\begin{align}
    M_D = \sqrt{M_N} O \sqrt{s} E O^\dagger \sqrt{M_N^*}\;.
\end{align}
Notice that $O$, $s$ and $E$ depend on $M_\nu$ and $M_N$, the neutrino Dirac mass matrix, we can calculate $M_D$  once the light and heavy neutrino masses and mixings are known.

Although the above method is applied to the general seesaw relation in the Hermitian case (cf. Eq.~\eqref{eq:seesaw}), no general $M_D$ could be obtained since $M_N$ is arbitrary~\cite{Senjanovic:2018xtu}.
In terms of the physical masses and neutrino mixing matrices, 
\begin{align}
\label{eq:Mv-MN}
    M_\nu = V_L^* m_\nu V_L^\dagger\;,\quad
    M_N = V_R m_N V_R^T\;,
\end{align}
thus we should have a priori knowledge of $V_R$ and $m_N$ besides the inputs of $m_\nu$ and $V_L = U_{\rm PMNS}^*$ with $U_{\rm PMNS}$ being the Pontecorvo-Maki-Nakagawa-Sakata (PMNS) matrix from the measurements of neutrino oscillation~\cite{Esteban:2020cvm}.

If $V_R = V_L$ is assumed,we could obtain~\cite{Senjanovic:2018xtu,Senjanovic:2016vxw}
\begin{align}
\label{eq:VLVR}
    M_D = V_L m_N \sqrt{\dfrac{v_L}{v_R} - \dfrac{m_\nu}{m_N}} V_L^\dagger\;.
\end{align}

While it is straightforward to calculate $M_D$ for a different $V_R$, the following condition 
\begin{align}
\label{eq:Hermiticity}
    \text{Im}\text{Tr}\left[\frac{v_L}{v_R}-\frac{1}{M_N}M_{\nu}^*\right]^n=0\;,\quad n=1,2,3\;
\end{align}
makes it more complicated, which results from the Hermiticity of $H$. The above relation implies that
the phases of light and heavy neutrino mass matrices are not independent~\cite{Senjanovic:2016vxw}.

That is to say, for any $V_R$ being assumed,  it is necessary to verify the condition in Eq.~\eqref{eq:Hermiticity} with the resulting heavy neutrino mass matrix $M_N$. Therefore, an appropriate choice of $V_R$ is crucial and non-trivial~\footnote{Very recently, an explicit closed form solution for the neutrino Dirac mass matrix in the MLRSM for the case $\mathcal{C}$ was obtained using the Cayley-Hamilton theorem~\cite{Kriewald:2024cgr}. }.

\subsection{parametrization of \tf{$V_R$}{VR} and \tf{$M_N$}{MN}}
\label{sec:choicesofVR}

Notice that if $v_L$ is real, the condition in Eq.~\eqref{eq:Hermiticity} is reduced to $\text{Im}\text{Tr}\left[M_N^{-1} M_{\nu}^*\right]^n=0$. This enables us to obtain possible forms of $M_N$ and $V_R$, the details of which are given in Appendix~\ref{app:Hermitcity}. 

We find that for Hermitian $M_D$ and real $v_L$ in the MLRSM for the case $\mathcal{P}$, the right-handed neutrino mixing matrix $V_R$ can be parameterized as
\begin{align}
\label{eq:new-VR}
    V_R=PV_L\sqrt{m_N m_\nu}^{-1}\;,
\end{align}
where $P$ is a Hermitian or anti-Hermitian matrix, 
\begin{align}
\label{eq:P-antiHermitian}
    P = \pm P^\dagger\;.
\end{align} 

For convenience, we can further write $V_R$ as
\begin{align}
\label{eq:VR-2}
    V_R = \hat{P} V_L\;, \quad \hat{P} \equiv P V_L \sqrt{m_N m_\nu}^{-1} V_L^\dagger\;.
\end{align}
Note that $P$ has the mass dimension one, while $\hat{P}$ is dimensionless.
As $V_L$ and $V_R$ are unitary~\cite{Senjanovic:2018xtu}, it follows that $\hat{P}$ must also be a unitary matrix, thereby imposing constraint on $P$.
If $V_R = V_L$, $\hat{P} = \mathds 1$, we readily get the Hermitian matrix $P = V_L \sqrt{m_N m_\nu} V_L^\dagger$.

From Eq.~\eqref{eq:new-VR}, one can construct the heavy neutrino mass matrix
\begin{align}
\label{eq:MN}
    M_N= PM_{\nu}^{-1} P^T\;,
\end{align}
which satisfies the condition in Eq.~\eqref{eq:Hermiticity}. 

If $V_R = V_L$, and $m_N = v_R/v_L m_\nu$, using Eq.~\eqref{eq:MN}, we can readily get $M_N = v_R/v_L M_\nu^*$. Thus the above parametrization of $M_N$ is compatible with the type-II seesaw dominance scenario.

\section{One-loop \tf{$\bar \theta$}{theta}}
\label{sec:theta}

As pointed out in Ref.~\cite{Kuchimanchi:2014ota}, $\bar\theta$ can be generated from the leptonic CP violation, which contributes to the Higgs potential at one-loop level:
\begin{align}
    V \supset \left[\alpha_2 \Tr(\Delta_R^\dagger \Delta_R) + {\rm h.c.}\right] \Tr (\tilde \Phi^\dagger \Phi)\;,
\end{align}
where the coupling $\alpha_2$ is complex and $\tilde{\Phi} \equiv \sigma_2 \Phi^* \sigma_2$.
It is shown that~\cite{Kuchimanchi:2014ota,Senjanovic:2020int}
\begin{align}
    \bar \theta_{loop} \simeq \dfrac{1}{16\pi^2} \dfrac{m_t}{m_b} \Im \text{Tr}\left( Y_R^\dagger Y_R [Y_1, Y_2] \right)\ln \dfrac{M_{Pl}}{v_R}\;,
\end{align}
where the Dirac Yukawa couplings $Y_R$ and $Y_{1,2}$ are defined in Eq.~\eqref{eq:lag-yuk}, and $M_{Pl}=1.22\times 10^{19}\gev$ denotes the Planck scale. In terms of the mass matrices, we have~\cite{Senjanovic:2020int}
\begin{align}
\label{eq:thetaloop}
    \bar \theta_{loop} &\simeq \dfrac{1}{16\pi^2} \dfrac{m_t}{m_b} \dfrac{1}{v_R^2 v^2 } \nn\\
    &\quad \times \Im \text{Tr}\left( M_N^T M_N^* [M_D, M_\ell] \right)\ln \dfrac{M_{Pl}}{v_R}\;.
\end{align}
where the charged lepton mass matrix $M_\ell$ is diagonal due to $U_e = \pm \mathds 1$.
If $V_R= V_L$,
by using the expressions of $M_N$ and $M_D$ given in Eqs.~\eqref{eq:Mv-MN}~\eqref{eq:VLVR}, we can easily verify that $\bar \theta_{loop}$ is exactly zero. This also applies when $V_R = \mathds 1$~\cite{deVries:2022nyh}.

\begin{figure}[h]
    \centering
    \includegraphics[width=0.8\linewidth]{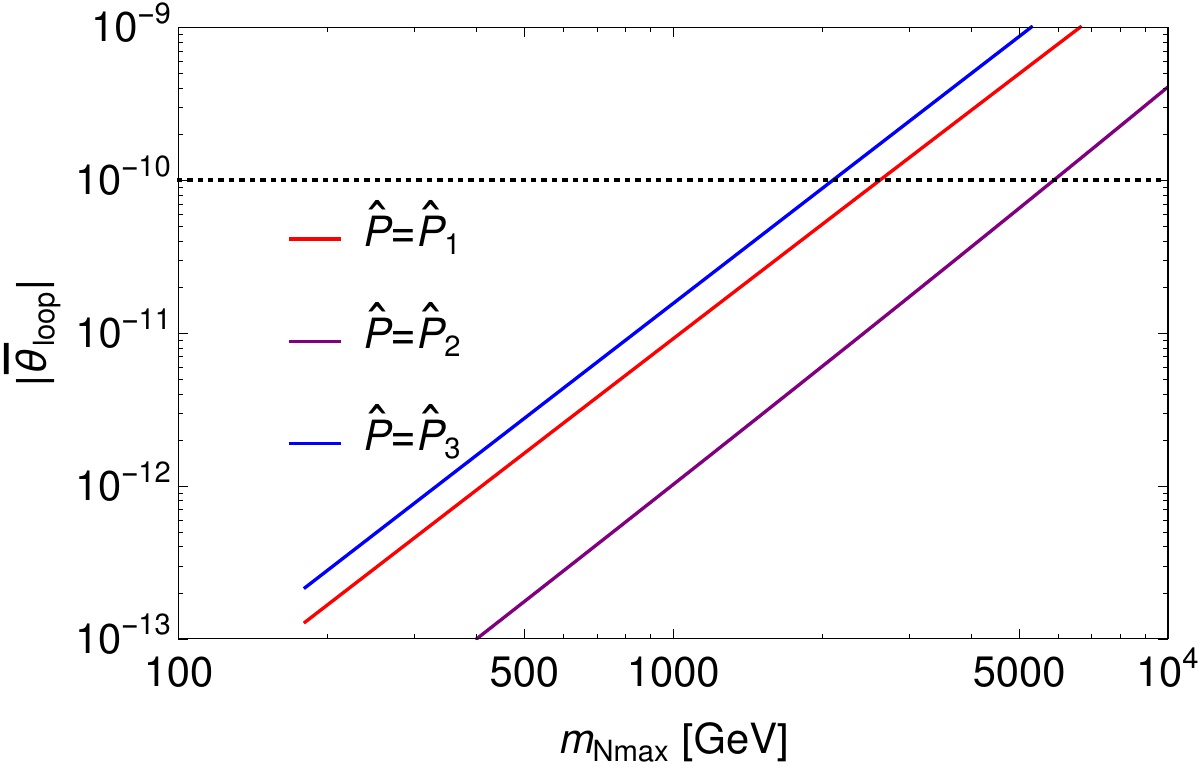}
    \caption{The magnitude of $\bar{\theta}_{loop}$ as a function of the heaviest sterile neutrino mass $m_{N{\rm max}}$, which is assumed to be $m_4$. 
    }
    \label{fig: theta term}
\end{figure}

In order to evaluate $\bar \theta_{loop}$ for other choices of $V_R$, we use the parametrization in Sec.~\ref{sec:choicesofVR}, and consider $V_R = \hat{P} V_L$ with the following textures of $\hat P$:
\begin{align}
\label{eq:cases-P} 
     \hat P_1&=i\begin{pmatrix}
        1\ &0\ &0\\
       0\ &0\ &1\\
        0\ &1\ &0
    \end{pmatrix}\;,\quad 
    \hat P_2=i\begin{pmatrix}
        0\ &1\ &0\\
       1\ &0\ &0\\
        0\ &0\ &1
    \end{pmatrix}\;,\quad
    \nonumber\\
    \hat P_3&=i\begin{pmatrix}
        \frac{1}{2}\ &\frac{1}{2}\ &-\frac{\sqrt{2}}{2}\\
        \frac{1}{2}\ &\frac{1}{2}\ &\frac{\sqrt{2}}{2}\\
        -\frac{\sqrt{2}}{2}\ &\frac{\sqrt{2}}{2}\ &0
    \end{pmatrix}\;.
\end{align}
One can directly verify that for these textures the matrix $P = \hat{P} V_L \sqrt{m_N m_\nu} V_L^\dagger$ is anti-Hermitian.   In Eq.~\eqref{eq:cases-P}, we have included the factor of $i$ to maintain the Hermiticity of the neutrino Dirac 
mass matrix $M_D$.

It is noted that in the type-I seesaw dominance scenario of the MLRSM for the case $\mathcal{P}$, where $v_L$ is negligibly small, $M_D$ in Eq.~\eqref{eq:VLVR} is anti-Hermitian if we assume $V_R = V_L$, a fact that appears to have been overlooked. 
We find that  $V_R = i V_L$, namely $\hat P = i\mathds{1}$, leads to $\bar\theta_{loop}=0$, which can be analytically verified~\footnote{In another study within the context of the left-right symmetric model with double seesaw~\cite{Patra:2023ltl}, $V_R = i V_L$ was considered. }.

We assume that the active neutrino masses $m_\nu \equiv \text{diag}(m_1,m_2,m_3)$ are in the normal hierarchy, and that the sterile neutrino masses  $m_N \equiv \text{diag}(m_4,m_5,m_6)$ are correlated with $m_\nu$:
\begin{align}
    m_1 m_4 = m_2 m_5 = m_3 m_6\;,
\end{align}
which is analogous to the assumption in Ref.~\cite{Patra:2023ltl}.
Furthermore, we choose the central values of the mixing parameters including the Dirac CP phase~\cite{Esteban:2020cvm} whereas the Majorana phases are set to be zero, and assume
\begin{align}
m_1 = 10^{-3}{~\rm eV}\;,\quad v_L = 1{~\rm eV}\;,\quad v_R = 15\tev\;.
\end{align}

For the textures in Eq.~\eqref{eq:cases-P}, we obtain the magnitude of non-vanishing $\bar\theta_{loop}$ as a function of the heaviest sterile neutrino $m_{\rm N max} = m_4$ in Fig.~\ref{fig: theta term}. 
Since $\bar\theta_{loop}$ approximately increases with the sterile neutrino masses $(m_{\rm Nmax })^{5/2}$, by requiring $|\bar \theta_{loop}|\lesssim 10^{-10}$ as an estimate~\footnote{We obtain that $\bar\theta_{loop}$ has definite sign for $\hat{P}$ being given in Eq.~\eqref{eq:cases-P}. 
Following Refs.~\cite{Kuchimanchi:2014ota,Senjanovic:2020int}, we take the bound $\mid \bar\theta_{loop}\mid \lesssim 10^{-10}$ without taking into account the tree-level contribution to $\bar\theta$. This can be achieved by making the {\it ad hoc} assumption that $\alpha_2$ is real at tree level~\cite{Kuchimanchi:2014ota}. 
Note that $\bar \theta= \arg \det (M_u M_d)$ is proportional to the overall imaginary part of $\alpha_2$~\cite{Senjanovic:2020int}.
If tree-level $\text{Im}~ \alpha_2$ is much larger than the one-loop value generated from leptonic CP violation, the upper limits on the sterile neutrino masses would be more stringent than those obtained using $\mid \bar\theta_{loop}\mid \lesssim 10^{-10}$. If, however, the tree-level and one-loop contributions are comparable and have opposite signs, their cancellation would invalidate the quoted limits, potentially allowing for heavier sterile neutrinos.
}, we obtain the upper bound on the sterile neutrino masses, which was highlighted in Ref.~\cite{Senjanovic:2020int}. For $\hat{P}=\hat P_1$, $\hat{P}_2$ and $\hat{P}_3$, we obtain $m_{\rm N max} \lesssim 2.5\tev$, $6\tev$ and $2\tev$, respectively. The sterile neutrinos with mass up to approximately $ 3\tev$ are within reach of direct searches at the LHC~\cite{ATLAS:2023cjo}, while the light sterile neutrinos with masses in the MeV-GeV range could make significant contributions to $0\nu\beta\beta$ decay~\cite{deVries:2022nyh}.

On one hand, $\bar\theta_{loop}$ varies with different choices of $\hat P$. On the other hand,
we emphasize that a small deviation from the chosen unitary matrix $V_R$ in our parametrization would not lead to an unacceptably large value of $\bar\theta_{loop}$. This is because the smallness of $\bar\theta_{loop}$ is attributed to the loop factor $1/(16\pi^2)$ as well as the suppression by the ratios $ M_N^2/v_R^2 $ and $M_D M_\ell/v^2$, as indicated in Eq.~\eqref{eq:thetaloop}.

As a benchmark, we take $\hat{P} = \hat{P}_1$ and assume
$m_4 =2.86 {~\rm TeV}$, $m_5=3.32 {~\rm GeV}$, and $m_6 =57.2 {~\rm MeV}$.
The heavy neutrino mass matrix is given by
\begin{widetext}
\begin{align}
    M_N=
    \left(\begin{array}{ccc}
        -1.95\times 10^{12}-7.05\times10^{5} i & -1.16\times 10^{12}+5.60\times 10^{10} i & 6.38\times 10^{11}+6.47\times 10^{10} i \\
        -1.16\times 10^{12}+5.60\times 10^{10} i & -6.96\times 10^{11}+6.69\times 10^{10} i & 3.85\times 10^{11}+2.03\times 10^{10} i \\
        6.38\times 10^{11}+6.47\times 10^{10} i & 3.85\times 10^{11}+2.03\times 10^{10} i & -2.09\times 10^{11}-4.24\times 10^{10} i \\
    \end{array}\right){~\rm eV}\;.
    \notag
\end{align}
Using the Senjanovic-Tello method, we obtain the matrices in Eq.~\eqref{eq:H-matrix2}
\begin{equation}
    \notag
    \begin{split}
        &O=
        \left(\begin{array}{ccc}
            -0.1344+0.04691i
            & -0.4861-0.006028i
            & 0.8648+0.003902i
            \\
            0.6396-0.0002750i
            & 0.6240+0.01683i
            & 0.4499-0.02296i
            \\
            0.7584+0.008545i
            & -0.6125+0.02193i
            & -0.2263-0.03073i
        \end{array}\right)\;,
        \\
        &E=\mathds{1}\;,\ s=
        \left(\begin{array}{ccc}
            8.576\times10^{-7}
            & 0
            & 0
            \\
            0
            & 4.602\times10^{-10}
            & 0
            \\
            0
            & 0
            & 6.867\times10^{-13}
        \end{array}\right)\;,
    \end{split}
\end{equation}
and the neutrino Dirac mass matrix
\begin{equation}
    M_D=
    \left(\begin{array}{ccc}
        545380. & 343623.+13636. i & -204348.+16763. i 
        \\
        343623.-13636. i & 272102. & -116364.+17374. i 
        \\
        -204348.-16763. i & -116364.-17374. i & 109404. \\
    \end{array}\right){~\rm eV}\;,
    \notag
\end{equation}
\end{widetext}
and the resulting value of $\bar{\theta}_{loop}$ is $-1.241335\times10^{-10}$.

It is worth noting that due to $ s_\alpha t_{2\beta} \propto \Im \alpha_2$~\cite{Kriewald:2024cgr}, the neutrino Dirac mass matrix $M_D$ is not exactly Hermitian. 
A delicate examination of non-Hermitian $M_D$ was recently conducted in Ref.~\cite{Kiers:2022cyc}. Nevertheless, the correlation between $M_D$ and $\bar \theta$  poses a challenge for recursive evaluation. Unless there is accidental cancellation, assuming that $M_D$ is exactly Hermitian, as we have done, is adequate for estimating the upper limit on the sterile neutrino masses.

\section{Conclusion}
\label{sec:conclusion}

In this work, we have proposed a parametrization of right-handed neutrino mixing $V_R = P V_L \sqrt{m_N m_\nu}^{-1} $ with $P$ being a Hermitian or anti-Hermitian matrix in the MLRSM for the case $\mathcal{P}$, and constructed heavy neutrino mass matrix as $M_N = P M_\nu^{-1} P^T$. In this parametrization, the Hermiticity of the neutrino Dirac mass matrix $M_D$ is maintained.

We then evaluate the one-loop $\bar\theta$ generated from leptonic CP violation for the general seesaw relation with explicit examples of $V_R$ and obtain non-vanishing $\bar \theta_{loop}$ as a function of the sterile neutrino masses. By requiring $|\bar \theta_{loop}| \lesssim 10^{-10}$, we obtain the upper bound on the sterile neutrino masses.

Our parametrization of $V_R$ and $M_N$ is pertinent for other phenomenological investigations. Specifically, it remains to explore the impact of the texture of $V_R$, or equivalently $\hat{P}$, on the LHC searches and $0\nu\beta\beta$ decay. Their interplay might be able to disentangle different $V_R$'s beyond the prevailing choices currently adopted in the literature.

\appendix

\section{Hermitcity of \tf{$H$}{H}}
\label{app:Hermitcity}

In Sec.~\ref{sec:choicesofVR}, we have provided the parametrization of the right-handed neutrino mixing $V_R$ and heavy neutrino mass matrix $M_N$, which guarantees the Hermiticity of the matrix $H$ hence neutrino Dirac mass matrix $M_D$. In this appendix, we will give more details.

From Eq.~\eqref{eq:Hermiticity}, we expand
\begin{align}
    &\left[\frac{v_L}{v_R}-\frac{1}{M_N}M_{\nu}^*\right]^n\nonumber\\
    =&C_1 + C_2\frac{1}{M_N}M_\nu^* +...+C_n(\frac{1}{M_N}M_\nu^*)^n\;,
\end{align}
where all the $C_n$ for $n\in\mathds{N}$ are real numbers. So all we need to check is $\left(M_N^{-1}M_{\nu}^*\right)^n$.

First, we assume
\begin{align}
\label{eq:MN-PQ}
    M_N = P M_\nu^{-1} Q\;,
\end{align}
where $Q$ is a matrix with dimension one. 

We observe that if $Q = \pm P^*$, 
\begin{align}
    &\text{Im}\text{Tr}\left[\left(\frac{1}{M_N}M_{\nu}^*\right)^{n}\right]
    \nonumber\\
    =&\pm\text{Im}\text{Tr}\left[(P^{-1*}M_{\nu}P^{-1}M_{\nu^*})^n\right]\;.
\end{align}
Defining $A = P^{-1^*} M_\nu$, we have
\begin{align}
    \Tr[ (A A^*)^n ]& = \Tr[ A^* (A A^*)^{n-1} A ]\nn\\
    &=\Tr[ (A^* A)^n ]\;,
\end{align}
so that 
\begin{align}
    \text{Im}\text{Tr}\left[\left(\frac{1}{M_N}M_{\nu}^*\right)^{n}\right] 
    = 
    \text{Im}\text{Tr}\left[\left(\frac{1}{M_N}M_{\nu}^*\right)^{n*}\right]\;,
\end{align}
which implies that
\begin{align}
    \text{Im}\text{Tr}\left[\left(\frac{1}{M_N}M_{\nu}^*\right)^n\right] = 0\;.
\end{align}
Therefore, the condition in Eq.~\eqref{eq:Hermiticity} is satisfied. 

From Eq.~\eqref{eq:MN-PQ}, we have
\begin{align}
    M_N
     \label{eq:MN2}
    =&PV_L m_{\nu}^{-1}V_L^T Q\nonumber\\
    =& V_R m_N V_R^T\;.
\end{align}
To find a possible form of $V_R$, we define
\begin{align}
    F = \sqrt{m_\nu} X \sqrt{m_N}\;,
\end{align}
where $X$ is an orthogonal matrix, $X X^T =\mathds 1$,
and obtain
\begin{align}
\label{eq:F-decompose}
    m_{\nu}=Fm_N^{-1} F^T\;.
\end{align}

Then Eq.~\eqref{eq:MN2} becomes
\begin{align}
\label{eq:MN-solve}
    PV_L (F^T)^{-1} m_{N}F^{-1}V_L^T Q = V_R m_N V_R^T\;.
\end{align}
If 
\begin{align}
 \label{eq:ture construction}
    V_R=PV_L (F^T)^{-1}\;,\quad
    V_R^T=F^{-1}V_L^T Q\;,
    \end{align}
the relation in Eq.~\eqref{eq:MN-solve} must be satisfied. Hence, $Q = P^T = \pm P^*$ with $P$ being a Hermitian or anti-Hermitian matrix.
Without loss of generality, we assume $X=\mathds 1$, and then $F=\sqrt{m_\nu m_N} $. Other choices of $X$ can also yield appropriate $V_R$ and $P$, which may also be of interest.

\section*{Acknowledgements}
GL would like to thank Jordy de Vries, Miha Nemev\v{s}ek, Fabrizio Nesti, and Juan Carlos Vasquez for illuminating discussions, and Ravi Kuchimanchi for helpful correspondence regarding Ref.~\cite{Kuchimanchi:2014ota}, and Chengcheng Han for insightful comments. DYL is grateful to Ken Kiers for generously sharing the codes of Ref.~\cite{Kiers:2022cyc} and for providing valuable suggestions and help. 
This work is partly supported by the National Natural Science Foundation of China under Grant No. 12347105, the Guangdong Basic and Applied Basic Research Foundation (2024A1515012668), and the Fundamental Research Funds for the Central Universities, Sun Yat-sen University (23qnpy62), and SYSU startup funding.

\setcounter{equation}{0}
\renewcommand\theequation{A.\arabic{equation}}

\bibliographystyle{apsrev4-1}
\bibliography{reference}
\end{document}